\documentclass[a4paper,12pt]{article}

\usepackage[square]{natbib}

\usepackage{marvosym}
\usepackage{tikzsymbols}
\usepackage{booktabs}
\usepackage{xspace,mathtools}

\usepackage{ctable}

\usepackage[lined,titlenumbered]{algorithm2e}

\usepackage{amssymb}
\usepackage{amsmath}
\usepackage{color}

\usepackage{amsthm}

\usepackage{multirow}

\usepackage{url}

\newtheorem{lemma}{{\sc Lemma}}

\newtheorem{theorem}{{\sc Theorem}}

\newtheorem{example}{Example}

\newtheorem{definition}{Definition}

\begin{document}

	\sloppy

\title{The Maximin Support Method: An Extension of the D'Hondt Method
  to Approval-Based Multiwinner Elections}

\author{Luis S\'anchez-Fern\'andez\\ 
Dept. Telematic Engineering,\\ Universidad Carlos III de
    Madrid,\\ E-28911 Legan\'es, Spain
\and
Norberto Fern\'andez Garc\'{\i}a\\
Centro Universitario de la Defensa,\\ Escuela Naval
  Militar,\\ E-36920 Mar\'{\i}n, Spain
\and
Jes\'us A. Fisteus\\
Dept. Telematic Engineering,\\ Universidad Carlos III de
Madrid,\\ E-28911 Legan\'es, Spain
\and
Markus Brill\\
Institute of Software Engineering and Theoretical
  Computer Science,\\ Technische Universit\"at Berlin,\\
  Ernst-Reuter-Platz 7, D-10587 Berlin, Germany
}

\maketitle

\begin{abstract}
We propose the maximin support method, a novel extension of the D'Hondt apportionment method to approval-based multiwinner elections. The maximin support method is based on maximizing the support of the least supported elected candidate. It can be computed efficiently and satisfies (adjusted versions of) the main properties of the original D'Hondt method: house monotonicity, population monotonicity, and proportional representation. We also establish a close relationship between the maximin support method and Phragm\'{e}n's voting rules. 
\end{abstract}

\newcommand{\MB}[1]{{\color{blue}\textbf{MB: #1}}}

\newcommand{\ms}{maximin support\xspace}
\newcommand{\Ms}{Maximin support\xspace}
\newcommand{\MS}{Maximin Support\xspace}
\newcommand{\mm}{\mathit{maximin}}
\newcommand{\mms}{\mathit{MMS}}


\section{Introduction}
\label{sec:intro}

Decision making based on the aggregation of possibly conflicting
preferences is a central problem in the field of social choice and
has received a considerable amount of attention from the artificial
intelligence research community~\citep{conitzer:cacm,
  brandt2016handbook, votingBookDAI, Skowron201567, aziz:scw,
  elkind:scw17, elkind:ijcai11, betzler:jair13}. A voting system takes as input the preferences of agents over a set of candidates, and outputs one or several candidates as the collective choice. We study \emph{multiwinner elections}, where a subset of candidates of a fixed size needs to be selected.

Multiwinner elections are often used in scenarios in which it
is desirable that the set of selected candidates {\it represents} different opinions or preferences of the electorate. For instance, this
is the case in parliamentary elections.

Representative multiwinner voting rules can also be applied in
multi-agent systems. For instance, consider the scenario in which a
group of friends usually goes to the cinema one or several times a
week.\footnote{A similar scenario has been discussed by
  \citet{skowron2016finding} and by \citet{elkind:scw17}.} Suppose
that this group of friends selects the set of movies that they are
going to watch within a certain time period (for instance, a month)
using a multiwinner voting rule.\footnote{The use of multiwinner
  voting rules for preference aggregation in recommender systems has
  also been proposed by \citet{naamani2014preference}.} In this
scenario, if a minority of the members of the group have different
tastes from the majority, the minority may demand that a proportion of
the movies (proportional to the size of the minority) is selected from
the movies the minority likes.

Another, completely different, example is the selection of teachers
for a school.\footnote{This example was suggested by Vincent
  Conitzer.} Suppose that a school has to hire a number of
teachers. All the classes in the school have the same size. The school
director decides to run an election to select the teachers, in which
parents have to give their preferences with respect to the candidates
that have applied for the teacher positions.\footnote{We assume that
  the election is run among all the parents whose children belong to
  the same grade.} In this scenario, the ideal situation is to select
and assign teachers in such a way that all the parents like the
teacher assigned to their children’s class. Therefore, it does not
matter if a teacher is only liked by a small minority of the parents,
as long as this minority consists of the parents of the children that
are assigned to this teacher.

Other examples of scenarios in which it is necessary to select a
number of candidates or choices that are representative of the
preferences of a group of agents have been discussed
by~\cite{skowron2016finding}, by~\cite{lu2011budgeted}, and
by~\cite{conitzer2017fair}.

One of the basic characteristics of a multiwinner voting rule is the
way in which agents (voters) cast their votes. Two alternatives are
commonly used: 1) ranked ballots, in which agents have to provide a
total order of the candidates; and 2) approval ballots, in which
agents simply approve as many candidates as they like. One of the
advantages of approval ballots compared to ranked ballots is the
simplicity of the ballots~\citep{laslierApproval}.

In this study, we propose an approval-based multiwinner voting rule
that can be used in scenarios in which it is necessary to select a
representative set of winners. Our voting rule, which we call the \emph{\ms method}, is an extension of the
well known D'Hondt method of apportionment (in the USA, the latter is also known as Jefferson's method) to approval-based multiwinner elections. The \ms method is
inspired by the defining property of the D'Hondt method: D'Hondt selects a set of
winners that maximizes the support of the least supported winner (a
detailed explanation of this concept will be provided later). 

Together with {\it seq-Phragm\'en},\footnote{seq-Phragm\'en was proposed by the Swedish mathematician Lars Edvard Phragm\'en~\citep{phragmen:p1, phragmen:p2, phragmen:pOpt, phragmen:p4} in the 19th century. In simultaneous and independent work to ours, \citet{brill:phragmen} proved that seq-Phragm\'en satisfies PJR and can be computed in polynomial time.} 
the \ms method is the first polynomial-time computable methods that is known to satisfy {\it proportional
  justified representation (PJR)}, a representation axiom proposed by \citet{pjr-aaai}. Moreover, like the
D'Hondt method, the \ms method also satisfies (adjusted versions of) house monotonicity and population monotonicity.

The remainder of this paper is organized as
follows. Section~\ref{sec:not} provides a summary of the main
notations and symbols used throughout the paper. Section~\ref{sec:dhondt} reviews the D'Hondt
method of apportionment, and Section~\ref{sec:support} introduces the concept of support distributions, on which the new method is based. The \ms method is defined in Section~\ref{sec:odh} and its axiomatic properties are analyzed in Section~\ref{sec:odhprop}.
Section~\ref{sec:comp} compares the \ms method with other extensions of the D'Hondt method, and Section~\ref{sec:concl} concludes. 

\section{Preliminaries}
\label{sec:not}

Let $C$ be a finite set of \emph{candidates} and $N = \{1, \dots, n\}$ be a set of $n$ \emph{voters}. Furthermore, $k$ denotes the number of winners to be selected. We assume $1 \le k \le |C|$ and $n \geq 1$.

For each $i \in N$, we let $A_i \subseteq C$ denote the \emph{approval ballot} of voter $i$. That is, $A_i$ is the subset of candidates that voter $i$ approves of.
A \emph{ballot profile} is a list $\mathcal{A}= (A_1, \dots, A_n)$ of approval ballots, one for each voter $i \in N$. 
An \emph{approval-based multiwinner election} can thus be represented by a tuple $(N, C, \mathcal{A}, k)$. 

An \emph{(approval-based multiwinner) voting rule} $R$ is a function that maps an election $\sigma=(N, C, \mathcal{A}, k)$ to a set $R(\sigma) \subseteq C$ of  $|R(\sigma)|=k$ candidates, interpreted as the \emph{winners} of $\sigma$ according to rule~$R$. 
During the execution of a voting rule, ties between candidates can occur. We assume that ties are broken using a fixed priority ordering over the universe of all possible candidates. For example, the priority ordering could be the lexicographic order, and this is the one we use in our examples.

An important subdomain of approval-based multiwinner elections is
defined by \emph{party-list elections}, where the set of candidates is
partitioned into parties and voters can vote for exactly one
party. Formally, a party-list election satisfies $C = P_1 \cup P_2
\cup \ldots \cup P_p$ and every approval ballot $A_i$ coincides with
one party list $P_j$. The ballot profile for a party-list election can
be summarized\footnote{We assume that the voting rules that we are
  going to consider are anonymous.} by a \emph{vote vector} $V=(v_1,
v_2, \ldots, v_p)$, where $v_j$ is the number of votes for party $P_j$
(i.e., $v_j= |\{i \in N: A_i=P_j\}|$). An \emph{apportionment method}
takes as input a vote vector $V=(v_1, v_2, \ldots, v_p)$ and a natural
number $k$ and outputs a \emph{seat distribution} $x=(x_1, \ldots,
x_p) \in \mathbb{N}_0^p$ with $\sum_{j=1}^p x_i = k$. The
interpretation is that party $P_j$ is allocated $x_j$
seats. Apportionment methods have been extensively studied in the
literature \citep{balYoung,Puke14a}.  Since the party-list setting is
a special case of the general approval-based multiwinner setting,
every approval-based multiwinner rule induces an apportionment method
\citep{brillmultiwinner}. An approval-based multiwinner rule is called an
\emph{extension} of an apportionment method if it induces it.
In this paper, we will introduce a novel extension of the apportionment method due to D'Hondt. 

\section{Review of the D'Hondt Apportionment Method}
\label{sec:dhondt}

The \emph{D'Hondt method} (aka \emph{Jefferson method}) is a particular example from a family of apportionment methods known as \emph{divisor methods}~\citep{electSyst,balYoung,Puke14a}.  
These methods assign seats to parties based on a sequence of \emph{divisors} $(d_1, d_2, d_3, \ldots)$, and different divisor methods differ in their choice of this sequence. 
Divisor methods can be illustrated by constructing a table in which columns correspond to parties and rows correspond to divisors. The entry in row $i$ and column $j$ is given by $v_j/d_i$, i.e., the number of votes of party $P_j$ divided by the $i$-th divisor. The divisor method then assigns the $k$ seats to the parties corresponding to the $k$ highest quotients in this table.

The D'Hondt method is defined via the divisor sequence $(d_1, d_2, d_3, \ldots) = (1,2,3,\dots)$. An example of the use of the D'Hondt method is shown in
Table~\ref{tab:dhondtEx1}, where five seats must be assigned to three
parties: $P_1$ (composed of candidates $a_1, \ldots, a_5$), $P_2$
(composed of candidates $b_1, \ldots, b_5$), and $P_3$ (composed of
candidates $c_1, \ldots, c_5$).
As shown in Table~\ref{tab:dhondtEx1}, the
D'Hondt method assigns three seats to party $P_1$, one to party $P_2$, and one to party $P_3$.
Using the lexicographic order to break ties among candidates of the same party, the set of elected candidates is $\{a_1, a_2, a_3, b_1, c_1\}$.

\begin{table}[htb]
\centering
\begin{tabular}{lrrrr} \toprule
\multicolumn{2}{l}{\bf Parties} & 
$P_1$ & $P_2$ & $P_3$ \\ \midrule

\multicolumn{2}{l}{\bf Votes ($v_j$)} & 
$5\thinspace 100$ & $3\thinspace 150$ & $1\thinspace 750$ \\ \midrule

\multirow{5}{*}{\bf Divisors} & $d_1=1$ &
${\bf 5\thinspace  100.0}$ & ${\bf 3\thinspace  150.0}$ & ${\bf 1\thinspace  750.0}$ \\ 

 & $d_2=2$ & ${\bf 2\thinspace  550.0}$ & $1\thinspace  575.0$ & $875.0$ \\ 
 & $d_3=3$ & ${\bf 1\thinspace  700.0}$ & $1\thinspace  050.0$ & $583.3$ \\ 
 & $d_4=4$ & $1\thinspace  275.0$ & $787.5$ & $437.5$ \\ 
& $d_5=5$ & $1\thinspace  020.0$ & $630.0$ & $350.0$ \\ \midrule

\multicolumn{2}{l}{\bf Seats ($x_j$)} & 
$3$ & $1$ & $1$ \\ \bottomrule
\end{tabular}

\caption{Example of the use of the D'Hondt method. The $k=5$ highest quotients are marked in bold and correspond to the seat distribution $(3,1,1)$.}
\label{tab:dhondtEx1}
\end{table}

An important proportionality axiom for apportionment methods is \emph{lower quota}, which requires that each party $P_j$ is allocated \emph{at least} $\lfloor k \frac{v_j}{n} \rfloor$ seats. It is well known that the D'Hondt method is the only divisor method satisfying lower quota~\citep{electMath, balYoung}. Moreover, the D'Hondt method satisfies \emph{house monotonicity} and \emph{population monotonicity} (see Section~\ref{sec:odhprop} for definitions of these properties).

\section{A Formal Model for the Concept of Support}
\label{sec:support}

In this section, we formally introduce the notion of {\it support}, on which our extension of the D'Hondt method will be based. 

One way of viewing the D'Hondt apportionment method is to consider
that a vote received by a party is distributed among the elected
candidates from that party. For instance, in the example illustrated
in Table~\ref{tab:dhondtEx1}, each of the three elected candidates
from party $P_1$ is supported by $5100/3= 1700$ voters, the
elected candidate from list party $P_2$ is supported by $3150$
voters, and the elected candidate from party $P_3$ is supported by $1750$ voters. By definition, the D'Hondt method chooses a seat
distribution that maximizes the support of the least supported
candidate~\citep{electMath}. (For instance, assigning two seats to
party $P_2$ would mean that those two candidates have a support of
only $1575$ voters.)

We now generalize this notion to the setting of approval-based multiwinner elections, by distributing votes in the form of approval ballots among the elected candidates. A natural condition is that the vote of a voter can only be distributed among those candidates that are approved by that voter. In general, there may be many different ways of distributing votes, leading to different support values for candidates. 
We now present a formal model of support without fixing any particular way of
distributing the votes.

\newcommand{\supp}{\mathit{supp}}
\newcommand{\sd}{\mathcal{F}}
\newcommand{\sdo}{\mathcal{F}^{\textrm{opt}}}

For an approval-based multiwinner election $\sigma = (N, C, \mathcal{A}, k)$ and a nonempty subset {$D \subseteq C$} of candidates, we define the family $\sd_{\sigma,D}$ of {\it support
  distribution functions} as the set of all functions that distribute support among the candidates in $D$.
Formally, $\sd_{\sigma,D}$ consists of
all functions $f: (N \times D) \rightarrow [0,1]$ satisfying

\begin{eqnarray*}
	f(i,c) &=& 0  \quad\text{for all $i \in N$ and $c \in D \setminus A_i$, and} \\
	\sum_{c \in A_i \cap D} f(i,c) &=&1 \quad \text{for all $i \in N$ with $A_i \cap D \neq \emptyset$.} 
\end{eqnarray*}

For each voter $i \in N$, $f(i, c)$ is the fraction of voter $i$'s vote that is ``assigned'' to candidate $c$. Note that the definition requires that $f(i,c)=0$ whenever $c \notin A_i$. Thus, the vote of a voter is distributed only among those candidates that are approved by that voter. 
Given a support distribution function $f \in \sd_{\sigma,D}$ and a candidate $c \in D$, we let $\supp_f(c)$ denote the total support received by $c$ under $f$, i.e., $\supp_f(c)= \sum_{i \in N} f(i, c)$.

\begin{table}[htb]
\centering
\begin{tabular}{lr} \toprule
{\bf Approval ballot} & {\bf Votes} \\ \midrule
$\{c_1, c_2\}$ & $10\thinspace 000$ \\
$\{c_1, c_3\}$ & $6\thinspace 000$ \\ 
$\{c_2\}$ & $4\thinspace 000$ \\ 
$\{c_3\}$ & $5\thinspace 500$ \\ 
$\{c_4\}$ & $9\thinspace 500$ \\ 
$\{c_5, c_6, c_7\}$ & $5\thinspace 000$ \\ 
$\{c_5\}$ & $3\thinspace 000$ \\ \bottomrule
\end{tabular}
\caption{Ballot profile for election $\sigma_1$}
\label{tab:odhEx2}
\end{table}

\begin{example}\label{ex:1}
Consider the election $\sigma_1= (N, C, \mathcal{A}, k)$ with $k=3$ and $C=\{c_1,c_2,c_3,c_4,c_5,c_6,c_7\}$.  
There are $n=43000$ voters and the ballot profile is shown in Table~\ref{tab:odhEx2}.
Consider the subset $D=\{c_1,c_3,c_5\}$ and let $f$ be the (unique) function in $\sd_{\sigma_1,D}$ with $f(i,c_1)= 0.4$ for each voter $i$
  with $A_i= \{c_1,c_3\}$ 
  (thus $f(i,c_3) = 0.6$ for those voters). 
  Thus, $f$ assigns $2400$ out of the $6000$ $\{c_1,c_3\}$-votes to $c_1$ and the remaining $3600$ $\{c_1,c_3\}$-votes to $c_3$, resulting in the following support values.

\begin{eqnarray*}
\supp_{f}(c_1) &=& 
\sum_{\mathclap{i: A_i= \{c_1,c_2\}}} f(i, c_1) + \sum_{\mathclap{i: A_i= \{c_1, c_3\}}} f(i, c_1) = 10\thinspace 000 + 2\thinspace 400 = 12\thinspace 400 \\
\supp_{f}(c_3) &=&
\sum_{\mathclap{i: A_i= \{c_1, c_3\}}} f(i, c_3) +
\sum_{\mathclap{i: A_i= \{c_3\}}} f(i, c_3) =
3\thinspace 600 + 5\thinspace 500 = 9\thinspace 100 \\
\supp_{f}(c_5) &=&
\sum_{\mathclap{i: A_i= \{c_5, c_6, c_7\}}}
f(i, c_5) + \sum_{\mathclap{i: A_i= \{c_5\}}} f(i, c_5) =
5\thinspace 000 + 3\thinspace 000 = 8\thinspace 000
\end{eqnarray*}
\end{example}

\section{The \MS Method}
\label{sec:odh}

We now propose an extension of the D'Hondt method to approval-based multiwinner elections. It is based on the same principle as the D'Hondt method, in that the support for the least supported elected candidates should be as large as possible. We therefore refer to this novel method as \emph{\ms method} ($\mms$). 
The \ms method chooses candidates sequentially until the desired
number $k$ of candidates have been selected. In every iteration, a
candidate with the greatest support is chosen, under the condition
that only support distribution functions maximizing the support for
the least supported candidate are considered.

In order to formally define the method, we need the following notation. 
For an approval-based multiwinner election $\sigma =
(N, C, \mathcal{A}, k)$ and a nonempty candidate
subset $D \subseteq C$, let
$\mm(\sigma, D)$ denote the maximal
support for the least supported candidate in $D$,
where the maximum is taken over all support distribution functions in $\sd_{\sigma,D}$.
Formally,

\[
\mm(\sigma, D)= \displaystyle
\max_{f \in \sd_{\sigma,D}} \min_{c \in D} \supp_f(c)\text.
\]

Furthermore, we let 
$\sdo_{\sigma,D}$ denote the
nonempty\footnote{Since $\sd_{\sigma,D}$ may be an infinite set, we need to make sure that the function $\min_{c \in D} \supp_f(c)$ attains a maximum over this set. We will see in the proof of Theorem~\ref{th:complex} that the corresponding optimization problem can be formulated as a feasible and bounded linear program. It follows that $\sdo_{\sigma,D} \neq \emptyset$ and that $\max_{f \in \sd_{\sigma,D}} \min_{c \in D} \supp_f(c)$ indeed exists.} 
set of support distribution functions that maximize the
support for the least supported candidate in {$D$} for
election $\sigma$, i.e., 
\[
\sdo_{\sigma,D} = \{f \in \sd_{\sigma,D}:  \forall c \in D, 
\supp_f(c) \geq \mm(\sigma, D)\}.
\]
Functions in $\sdo_{\sigma,D}$ are called {\it optimal support distribution functions}.

We are now ready to present the \ms method. Given an approval-based multiwinner election $\sigma$, the set $W=\mms(\sigma)$ is determined by starting with $W=\emptyset$ and  iteratively adding candidates until $|W|=k$. 
In each iteration, we add to $W$ the unelected candidate receiving the greatest support, under the condition that only optimal support distributions functions are considered.\footnote{Restricting attention to optimal support distribution functions ensures that support for previously elected candidates is not ignored when searching for new support distribution function; see also Theorem~\ref{th:lastIsLeast}.}
More precisely, for each candidate $c \in C \setminus W$, we compute an optimal support distribution function $f_c$ for the set $W \cup \{c\}$ and determine the total support $\supp_{f_c}(c)$ that $c$ receives under $f_c$. The candidate maximizing this value is then added to $W$.
The procedure is formally described in Algorithm~\ref{tab:odh}. 

\begin{algorithm}[!htb]

\LinesNumbered

\TitleOfAlgo{\MS Method ($\mms$)}
\label{tab:odh}

\KwData{approval-based multiwinner election $\sigma = (N, C, \mathcal{A}, k)$}

\KwResult{subset $W \subseteq C$ of candidates with $|W|=k$}

\Begin{
\DontPrintSemicolon $W= \emptyset$\;
\For{$j$=1 {\bf to} $k$}{ 
\label{alg1:loop1} 
\ForEach{$c \in C \setminus W$}{ 
\label{alg1:loop2} 
\DontPrintSemicolon compute $f_c \in \sdo_{\sigma,W \cup \{c\}}$\;
\label{alg1:fc1}
        \PrintSemicolon
        $s_c= \supp_{f_c}(c)$
\label{alg1:sc}
      }
\label{alg1:endloop2} 
\DontPrintSemicolon $w = {\displaystyle \arg \max_{c \in C \setminus W} s_c}$\;
\label{alg1:ci} 
$W= W \cup \{w\}$} 
\label{alg1:endloop1} 
\Return{$W$}
}
\end{algorithm}

Since the set $\sdo_{\sigma,W \cup \{c\}}$ of optimal support distribution functions may contain more than one function, the value of $s_c= \supp_{f_c}(c)$ could potentially depend on the choice of $f_c \in \sdo_{\sigma,W \cup \{c\}}$. The following result implies that this is not the case.

\newcommand{\abe}{approval-based multiwinner election\xspace}

\begin{theorem}
\label{th:lastIsLeast}
Let $\sigma=(N, C, \mathcal{A}, k)$ be an \abe. The following holds for each 
$j \in \{0, \ldots, k-1\}$.   

Let {$W^j$} denote the set of the first $j$ candidates chosen by the \ms method when applied to $\sigma$.
Then, for each candidate $c \in C \setminus W^j$ and for each 
optimal support distribution function 
$f_c \in \sdo_{\sigma, (W^j \cup \{c\})}$,
\[\supp_{f_c}(c) = \mm(\sigma, W^j \cup \{c\})\text.\]
\end{theorem}

Theorem~\ref{th:lastIsLeast}, whose proof can be found in \ref{app:proof}, states that in every iteration the candidate $c$ added to $W$ is among the least supported candidates under every optimal support distribution function. The support of this candidate thus equals $\mm(\sigma, W \cup \{c\})$, which (by definition) is independent of the particular $f_c \in \sdo_{\sigma,W \cup \{c\}}$ that was chosen in line \ref{alg1:sc} of the algorithm.

This result gives rise to an interesting alternative formulation of the \ms method. In this equivalent formulation, there is no need to choose an optimal support distribution function from $\sdo_{\sigma,W \cup \{c\}}$; rather, $s_c$ is directly defined as $\mm(\sigma, W \cup \{c\})$. A natural interpretation of this definition is that the value $s_c$ measures the effect that the addition of a potential candidate would have on the maximal support for the least supported candidate.

The next theorem establishes that the \ms method can be computed efficiently.

\begin{theorem}
\label{th:complex}
The \ms method can be computed in polynomial time. 
\end{theorem}

\begin{proof}
	It is sufficient to show that, for any subset $D\subseteq C$ of candidates, an optimal support distribution function $f \in \sdo_{\sigma, D}$ can be computed in polynomial time. For a given approval-based multiwinner election $\sigma = (N, C, \mathcal{A}, k)$ and a $D \subseteq C$, consider the following linear program, containing a variable $f(i,c)$ for each $i \in N$ and $c \in A_i \cap D$, and an additional variable $s$.
	
\begin{alignat*}{3}
 & \text{maximize} &s & \\
 & \text{subject to} \quad& 
    \sum_{\mathclap{{i \in N: c \in A_i \cap D}}} f(i,c) & \ge s, &\text{for all $c \in D$}\\
	&& \sum_{\mathclap{{c \in A_i  \cap D}}} f(i,c) & = 1, \quad &\text{for all $i \in N$ with $A_i \cap D \neq \emptyset$}\\
	&& f(i,c) &\ge 0, &\text{for all $i \in N$ and $c \in D$} \nonumber
\end{alignat*}

The first set of constraints require that the support for the least supported candidate in $D$ is at least $s$, while the remaining constraints ensure that the variables $f(i,c)$ encode a valid support distribution function.\footnote{Note that constraints of the form $f(i,c) \le 1$ are not necessary because each variable $f(i,c)$ is non-negative and appears in a constraint of the form $\sum_{{{c \in A_i  \cap D}}} f(i,c)  = 1$.} 
Therefore, optimal solutions of this linear program correspond to optimal support distribution functions.  
Since linear programming problems can be solved in polynomial time \citep{lpIsP}, this concludes the proof. 
\end{proof}

We conclude this section by illustrating the \ms method with
an example.

\begin{example}
	Consider again the election $\sigma_1$ from Example~\ref{ex:1}.
In the first step ($j=1$), the value $s_c=\mm(\sigma_1,\{c\})$ equals the approval score of candidate $c$, i.e., $s_c = |\{i \in N: c \in A_i\}|$ for all $c \in C$. Therefore, the approval winner ${c_1}$ (with $s_{c_1}=16000$) is chosen. The corresponding support distribution function $f$ satisfies $f(i,{c_1})=1$ for all $i \in N$ with $c_1 \in A_i$.

In the second step ($j=2$), we have $W=\{{c_1}\}$ and we need to compute the value $s_x = \mm(\sigma_1,\{{c_1},x\})$ for every $x \in C\setminus\{{c_1}\}$. 
For example, for candidate ${c_2}$ we get $s_{c_2} = \mm(\sigma_1,\{{c_1},{c_2}\}) = 10000$; the corresponding support distribution function~$f$ assigns $4000$ out of the $10000$ $\{{c_1},{c_2}\}$-votes to~${c_1}$ and the remaining $6000$ to ${c_2}$. 
A better value is achieved by candidate~${c_3}$. The support distribution realizing $s_{c_3} = \mm(\sigma_1,\{{c_1},{c_3}\}) = 10750$ assigns all $10000$ $\{{c_1},{c_2}\}$-votes to~${c_1}$, all $5500$ $\{{c_3}\}$-votes to~${c_3}$, and divides the $6000$ $\{{c_1},{c_3}\}$-votes between~${c_1}$ and~${c_3}$ such that both candidates have a total support of $10750$ each. 
Computing the other values, we get $s_{c_4}=9500$, $s_{c_5}=8000$, and $s_{c_6}=s_{c_7}=5000$. Therefore, ${c_3}$ is selected as the second candidate.

In the third step ($j=3$), we have $W=\{{c_1},{c_3}\}$ and we need to compute the value $s_x = \mm(\sigma_1,\{{c_1},{c_3},x\})$ for every $x \in C\setminus\{{c_1},{c_3}\}$. 
It can be checked that 
$s_{c_2} = 8500$, 
$s_{c_4} = 9500$,
$s_{c_5} = 8000$, and
$s_{c_6} = s_{c_7} = 5000$.
Thus, candidate ${c_4}$ is chosen. There are several support distribution functions $f \in \sdo_{\sigma_1,\{{c_1},{c_3},{c_4}\}}$ realizing $s_{c_4} = \mm(\sigma_1,\{{c_1},{c_3},{c_4}\}) = 9500$; each of them assigns all $9500$ $\{{c_4}\}$-votes to ${c_4}$ and distributes the $2500$ votes containing ${c_1}$ or ${c_3}$ in such a way that ${c_1}$ and ${c_3}$ have a total support of at least $9500$ each.

In summary, we have $\mms(\sigma_1)=\{{c_1},{c_3},{c_4}\}$.  
\end{example}

\section{Axiomatic Properties of the \MS Method}
\label{sec:odhprop}

In this section, we show that the \ms method is indeed an extension of the
  D'Hondt method, and that it satisfies (adjusted versions of) several important properties that the latter satisfies. In particular, we we show that the \ms method satisfies   
  \emph{house monotonicity}, 
  \emph{weak support monotonicity} (a variant of population monotonicity), 
  and \emph{proportional justified representation}. The latter property generalizes the notion of \emph{lower quota} to {\abe}s.

\subsection{D'Hondt Extension}

We first show that the \ms method coincides with the D'Hondt
method in the party-list domain. Our approach is similar
to that of \citet{brillmultiwinner}.

\begin{theorem}
\label{th:odh_prop1}
The \ms method is an extension of the D'Hondt method. 
\end{theorem}

\begin{proof}

Consider a party-list election $\sigma =
{ (N, C, \mathcal{A}, k)}$ with $C = P_1 \cup \ldots \cup P_p$ and vote vector $V=(v_1, v_2, \ldots, v_p)$ (i.e., $v_r= |\{i \in N: A_i=P_r\}|$). 
Let {$W^j$} be the set of
the first {$j$} candidates chosen by $\mms$. Let $c$ be a
candidate in {$C - W^j$} and let
{$P_r$} be the party to which $c$ belongs. In
Theorem~\ref{th:lastIsLeast} we proved that $c$ will always be in the
set of the least supported candidates when we maximize the support for
the least supported candidate in {$W^j \cup
  \{c\}$}. But $c$ is approved only by the voters that approve all the
candidates in {$P_r$}, and no candidate in
{$P_r$} is approved of by any other voter. This means
that maximizing the support of the least supported candidate in
{$W^j \cup \{c\}$} depends only on the number $v_r$
of voters approving {$P_r$} and on the number of
candidates of {$P_r$} that are in {$W^j \cup \{c\}$}.

Therefore, the support for the least supported candidate is maximized
if the total support $v_r$ of party {$P_r$} is
distributed uniformly among all the candidates in {$P_r$}
that are in {$W^j \cup \{c\}$} (any other
distribution of the votes to {$P_r$} would make one or
several candidates in {$(W^j \cup \{c\}) \cap P_r$}
receive less support), and
\begin{equation*}
\mm(\sigma, W^j \cup \{c\})=
\frac{v_r}{|(W^j \cup \{c\}) \cap P_r|}
\end{equation*}
This is exactly the same calculation the D'Hondt method performs for
selecting candidates, so both methods must assign the same number of
seats to each party. 
\end{proof}

\subsection{House Monotonicity}
\label{sec:house-mon}

House monotonicity requires that all selected candidates are still selected when the number $k$ of winners is increased.

\begin{definition}
	An approval-based multiwinner voting rule $R$ is \emph{house monotonic} if, for any pair of elections $\sigma= {(N, C, \mathcal{A}, k)}$ and {$\sigma'= (N, C, \mathcal{A}, k+1)$}, it holds that $R(\sigma) \subset R(\sigma')$.
\end{definition}

Since the \ms method selects winners iteratively, house monotonicity is trivially satisfied. 

\begin{theorem}
	The \ms method is house monotonic. 
\end{theorem}

\subsection{Support Monotonicity}
\label{sec:pop-mon}

The standard definition of population monotonicity requires that
additional support for a candidate does not harm that candidate. For
instance, the definition of candidate monotonicity given
by~\cite{elkind:scw17} when restricted to approval-based multi-winner
elections leads to this notion of monotonicity. A natural extension of
this idea has been proposed by~\citet{2017arXiv171004246S} by
considering what happens when the support of several of the winners is
increased. A first version of the axiom (referred by
S\'anchez-Fern\'andez and Fisteus as {\it weak} support monotonicity)
requires that, when the support of a subset of the winners is
increased, at least one of those candidates must remain in the winning
set.

\begin{definition}
	An approval-based multiwinner voting rule $R$ satisfies \emph{weak support monotonicity} if the following statements hold for all {\abe}s $\sigma = (N, C, \mathcal{A},
  k)$ and for all nonempty subsets $G \subseteq
R(\sigma)$ of winning candidates:

\begin{enumerate}
\item (weak support monotonicity without population increase) Let $i \in N$ be a voter with $A_i \cap G= \emptyset$ and 
  consider the election $\sigma' = (N, C, \mathcal{A}', k)$,
  where $A'_j = A_j$ for all $j \in N \setminus \{i\}$ and $A'_i = A_i \cup G$.    
  Then, $R(\sigma') \cap G \ne \emptyset$.

\item (weak support monotonicity with population increase) Consider
  the election $\sigma'' = (N \cup \{n+1\}, C, \mathcal{A}'', k)$,
  where $A''_j = A_j$ for all $j \in N$ and $A''_{n+1}= G$.  Then,
  $R(\sigma'') \cap G \ne \emptyset$.
\end{enumerate}
\end{definition}

We note that this definition reduces to the standard definition of
population monotonicity when $|G|=1$, and thus, despite its name, weak
support monotonicity is slightly stronger than the standard version
of population monotonicity.

\begin{theorem}
\label{theo:odh_pop}
 The \ms method is weak support monotonic.
\end{theorem}

\begin{proof}

First of all, we observe that for any nonempty candidate set $X$ that
is disjoint from $G$, the maximum support of the least supported
candidate in $X$ (when distributing the votes only between the
candidates in $X$) is the same for $\sigma$, {$\sigma'$}, and
{$\sigma''$}. This is because the changes made in {$\sigma'$} and
{$\sigma''$} do not affect how the votes can be distributed between
the candidates in~$X$.

Let {$r$} be the $\mms$ iteration in which the first
candidate from $G$ is elected in election $\sigma$ and let
{$c^* \in G$} be such candidate. {For $0 \le j \le k$, let
  $W^{j}$, $W_{\sigma'}^{j}$, and $W_{\sigma''}^{j}$ be the first $j$
  candidates chosen by $\mms$ for elections $\sigma$,
  {$\sigma'$}, and~{$\sigma''$}. (For $j=0$,
  we have $W^0 = W_{\sigma'}^{0} = W_{\sigma''}^{0}= \emptyset$.)}

If at least one candidate from $G$ is selected within the first~{$(r-1)$} iterations of the execution of
$\mms$ for election {$\sigma'$}
{(respectively, for election $\sigma''$)}, the statement of the theorem holds. Therefore, we assume that in the first $r-1$ iterations no candidate from $G$ is selected for election $\sigma'$ (respectively, for election $\sigma''$). In this case, at each iteration the candidate added to the set of winners will be the same for $\sigma$ and $\sigma'$ (respectively, $\sigma$ and $\sigma''$) because the computation of $\mm$ is done over sets of candidates disjoint from $G$.
Consequently, $W^{(r-1)}= W_{\sigma'}^{(r-1)}$ (respectively,
$W^{(r-1)}= W_{\sigma''}^{(r-1)}$). 

We are going to prove that the
candidate chosen at iteration~$r$ for election~$\sigma'$
(respectively, for election $\sigma''$) belongs to $G$. First,
we observe that since $W^{(r-1)} \cap G = \emptyset$, for each
candidate $c \in C \setminus (W^{(r-1)} \cup G)$ the maximum support
of the least supported candidate in $W^{(r-1)} \cup \{c\}$ is the same
for elections $\sigma$, $\sigma'$, and $\sigma''$. It is therefore
sufficient to prove that the maximum support of the least supported
candidate in $W^{(r-1)} \cup \{c^*\}= W^r$ for election~$\sigma'$
(respectively, for election $\sigma''$) is greater than or equal to
the maximum support of the least supported candidate in $W^r$ for
election $\sigma$. Further, it is enough to find a support
distribution function $g \in \sd_{\sigma', W^r}$ (respectively, a
support distribution function $h \in \sd_{\sigma'', W^r}$) such that
for each candidate $c$ in $W^r$ the support of $c$ under $g$
(respectively, the support of $c$ under $h$) is greater than or equal
to the maximum support of the least supported candidate in $W^r$ for
election $\sigma$.

Consider any optimal support distribution function
$f \in \sdo_{\sigma, W^r}$. For
  election $\sigma'$ we can define $g$ as follows. If $A_i \cap W^r \ne
  \emptyset$, then let $g(j,c)= f(j,c)$ for each voter $j \in N$ and
  each candidate $c \in W^r$. If voter $i$ does not approve any of the
  candidates in $W^r$ in election $\sigma$ (that is, if $A_i \cap W^r=
  \emptyset$), then for each candidate $c \in W^r$ we have $f(i,c)=
  0$. In that case we define $g(j,c)= f(j,c)$ for each voter $j \in
  N$, $j \ne i$, and each candidate $c \in W^r$, $g(i, c^*)= 1$, and
  $g(i, c)= 0$ for each candidate $c \in W^{(r-1)}$.

{For election $\sigma''$, let $h(j,c)= f(j,c)$ for
  each voter $j \in N$ and each candidate $c \in W^r$, $h(n+1,c^*)=
  1$, and $h(n+1, c)= 0$ for each candidate $c \in W^{(r-1)}$.}

  Clearly, each candidate in $W^r$ receives a support
  under $g$ and $h$ that is greater than or equal to the support that the same candidate receives under~$f$. Moreover, since $f \in \sdo_{\sigma, W^r}$, all candidates in $W^r$ receive a support under~$f$ that is greater than or equal to the support of the least supported candidate in $W^r$ for election $\sigma$.
\end{proof}

\citet{2017arXiv171004246S} also consider a stronger axiom 
called {\it strong support monotonicity} (with and without
population increase) that requires that, if the support of a subset $G$
of the winners is increased, \emph{all} candidates in $G$ must remain
in the set of winners. The following example shows that $\mms$ does not satisfy this stronger requirement.

\begin{example}
Consider the election $\sigma_2= (N, C, \mathcal{A}, k)$ with $k=6$ and $C=\{a, b, c_1, \ldots, c_5\}$. There are $18$ voters casting the
following ballots: $13$ voters approve of $\{c_1, \ldots, c_5\}$, $2$
voters approve of $\{a, b\}$, $2$ voters approve of $\{a\}$, and $1$
voter approves of $\{b\}$. For this election, we have $\mms(\sigma_2) = \{a, c_1, \ldots, c_5\}$ (candidate $a$ is elected in the fourth iteration). 
If a new voter enters the election and approves of precisely $\{a,
c_1, \ldots, c_5\}$, then the sets of winners outputted by $\mms$ is
$\{a, b, c_1, c_2, c_3, c_4\}$ (candidate $a$ is now elected in the
third iteration while candidate $b$ is elected in the last one). This
example proves that $\mms$ violates strong support monotonicity with
population increase.

To prove that $\mms$ violates strong support monotonicity without
population increase, we modify $\sigma_2$ by adding a new candidate $d$ and a new voter approving of $\{d\}$. Let $\sigma_3$ denote this modified election. It is easy to check that $\mms(\sigma_3) = \mms(\sigma_2) = \{a, c_1, \ldots, c_5\}$. 
If the new voter changes his approval set to $\{a,c_1, \ldots, c_5,d\}$, then the set of $\mms$ winners is again given by
$\{a, b, c_1, c_2, c_3, c_4\}$.
\end{example}

We stress that other axioms related to population monotonicity could
be defined in addition to the two cases that we are considering in
this paper. One example could be that one voter changes her vote and
no longer approves of a certain candidate $c$ that was in the set of
winners and a new voter enters the election and approves only
candidate $c$.

\subsection{Proportional Representation}
\label{sec:pjr}

Finally, we consider axiomatic properties concerning the proportional
representation of voter groups. In particular, we will consider two
axioms that have recently been proposed: \emph{proportional justified
  representation} (PJR)~\citep{pjr-aaai} and \emph{extended justified
  representation} (EJR)~\citep{aziz:scw}.  Both PJR and EJR are
generalizations of the lower quota axiom (see
Section~\ref{sec:dhondt}) to the general approval-based multiwinner
setting: if a voting rule satisfies PJR or EJR, then its induced
apportionment method satisfies lower quota \citep{brillmultiwinner}.

In order to define PJR and EJR, we need some terminology. Let $\sigma= (N, C, \mathcal{A}, k)$ be an \abe.  Given a positive integer $\ell\in \{1, \ldots, k\}$, we say that a subset $N^*\subseteq N$ of voters is \emph{$\ell$-cohesive} if
  $|N^*| \geq \ell \frac{n}{k}$ and $|\bigcap_{i \in N^*} A_i| \geq \ell$.  
  A subset $D \subseteq C$ of candidates %
  provides 
  \emph{proportional justified representation for $\sigma$ ($\sigma$-PJR)} 
  if for all $\ell\in\{1, \ldots, k\}$ and all $\ell$-cohesive subsets $N^* \subseteq N$, it holds that 
  \begin{equation}\label{eq:pjr}
  	|D \cap (\bigcup_{i \in N^*} A_i)| \ge \ell \text. 
  \end{equation}
  And $D$ provides 
  \emph{extended justified representation for $\sigma$ ($\sigma$-EJR)} 
  if for all $\ell\in\{1, \ldots, k\}$ and all $\ell$-cohesive subsets $N^* \subseteq N$,
  \begin{equation}\label{eq:ejr}
  	\text{there exists a voter $i \in N^*$ with } \left|A_i \cap D\right| \geq \ell \text.
  \end{equation}
   
\begin{definition}
  An approval-based multiwinner voting rule $R$ satisfies \emph{proportional justified representation (PJR)} (respectively, \emph{extended justified representation (EJR)}) if $R(\sigma)$ provides $\sigma$-PJR (respectively, $\sigma$-EJR) for every \abe $\sigma$.
\end{definition}

Since (\ref{eq:ejr}) implies (\ref{eq:pjr}), every rule satisfying EJR also satisfies PJR. 

\begin{theorem}
The \ms method satisfies proportional justified representation.
\end{theorem}

\begin{proof}
{For the sake of contradiction, suppose that there exists an election $\sigma= (N, C, \mathcal{A}, k)$ and an $\ell$-cohesive group $N^* \subseteq N$ such that for the set $W = \mms(\sigma)$ of winners
  output by the \ms method we have $|W \cap (\bigcup_{i \in N^*} A_i)| < \ell$. Thus, there are $x > k -
  \ell$ candidates in $W$ that are not approved of by any voter in
  $N^*$, and therefore, the support of some of these $x$ candidates
  (and the maximum support of the least supported candidate in $W$)
  has to be strictly less than $\frac{n}{k}$ (to see why, observe that
  $\frac{|N|-|N^*|}{x} \leq \frac{n - \ell \frac{n}{k}}{x} = n
  \frac{k-\ell}{kx} < n \frac{k - \ell}{k(k - \ell)}= \frac{n}{k}$).}

{By Theorem~\ref{th:lastIsLeast}, at each iteration
  the candidate that is added to the set of winners is one of the
  least supported when we maximize the support of the least
  supported candidate in the current set of winners. Therefore, at the
  last iterations of $\mms$ for election $\sigma$ the support of the
  candidate that we add to the set of winners when we maximize the
  support of the least supported candidate is strictly less than
  $\frac{n}{k}$ (this happens for sure at least in the last
  iteration).}

	Let $j$ be the first iteration of $\mms$ for election
  $\sigma$ such that the maximum support of the least supported
  candidate in $W$ is less than $\frac{n}{k}$ and let $c$ be the
  candidate elected in such iteration. Let $c^*$ be a candidate that
  is approved of by all the voters in $N^*$ and that does not belong
  to $W$ (such candidate exists because $|\bigcap_{i \in N^*} A_i|
  \geq \ell$ but $|W \cap (\bigcup_{i \in N^*} A_i)| < \ell$). Since
  all the voters in $N^*$ approve $c^*$ and there are at most $\ell
  -1$ candidates in $W$ that are approved by some voters in $N^*$, if
  we add candidate $c^*$ to the set of winners instead of candidate
  $c$ at iteration $j$, the support of $c^*$ when we maximize the
  support of the least supported candidate would be at least
  $\frac{|N^*|}{\ell}$ (observe that if some of the candidates in $W$
  that are approved by some voters in $N^*$ had a support greater than
  $\frac{|N^*|}{\ell}$ we could iteratively pick each of such
  candidates and give the surplus coming from voters in $N^*$ to
  $c^*$). Observe now that $\frac{|N^*|}{\ell} \geq \frac{\ell
    \frac{n}{k}}{\ell} = \frac{n}{k}$, and therefore candidate $c^*$
  would be elected ahead of candidate $c$ at iteration $j$, a
  contradiction. 
\end{proof}

The following example shows that the \ms method does not satisfy the stronger axiom EJR.

\begin{example} \label{ex:EJR}
  Consider election $\sigma_4 = (N, C, \mathcal{A}, k)$ with $k=4$ and 
  $C=\{{{a_1, a_2, a_3, c_1, c_2, c_3, c_4}}\}$. There are 16 voters casting the following ballots: $5$ voters
  approve of $\{{{a_1, c_1, c_2, c_3, c_4}}\}$, $4$ voters
  approve of $\{{{a_2, c_1, c_2, c_3, c_4}}\}$, $3$ voters
  approve of $\{{a_3, c_1, c_2, c_3, c_4}\}$, 
  $2$ voters approve of $\{a_1\}$, one voter approves of $\{{{a_2}}\}$, and one voter approves of $\{a_3\}$.

The set of winners according to the \ms method is given by
$\mms(\sigma_4) = \{c_1, a_1, a_2, a_3\}$ (selected in this
order). The $12$ voters whose approval set contains $\{{{c_1, c_2,
    c_3, c_4}}\}$ form a $3$-cohesive group, but none of these voters
approves at least $3$ candidates in the set of winners. Therefore,
$\mms(\sigma_4)$ fails to provide $\sigma_4$-EJR, which implies that
$\mms$ does not satisfy EJR.
\end{example}

\section{Comparison With Other D'Hondt Extensions}
\label{sec:comp}

The \ms method is not the only way to extend the D'Hondt method to {\abe}s. 
In this section we are going to compare the \ms method with other extensions of the D'Hondt method. 
In particular, we will consider two classes of rules, both of which originated in Scandinavia in the 1890s.  
  The first class of rules was proposed by the Swedish mathematician \emph{Lars Edvard Phragm\'en} \citep{phragmen:p1, phragmen:p2, phragmen:pOpt, phragmen:p4}
  and the second class of rules was proposed by the Danish polymath \emph{Thorvald N.\ Thiele}~\citep{thiele:pav_rav}. 
  We briefly review some axiomatic and computational properties of those rules and illustrate their differences with the \ms method. For an extensive treatment of Phragm\'en's and Thiele's rules and their properties, we refer to the survey by \citet{2016arXiv161108826J}.

\subsection{Phragm\'en's Rules}

\newcommand{\phrag}{Phragm\'{e}n\xspace}
\newcommand{\maxP}{max-\phrag}
\newcommand{\seqP}{seq-\phrag}

\phrag's methods can be described as \emph{load distribution} methods. Every selected candidate induces one unit of \emph{load}, and this load needs to be distributed among the voters that approve of that candidate. For example, if there are $6$ voters approving candidate $c$ and we decide to select this candidate for the committee, then one possible way of distributing the load would be to give a load of $\frac{1}6$ to each of those voters. However, it is not required that the load is distributed evenly among the approvers: different approvers of $c$ could be assigned different (non-negative) loads, as long as the loads associated with each selected candidate sum up to $1$. The goal is to choose a committee $W$ such that the load distribution is as \emph{balanced} as possible. Different interpretations of “balancedness” lead to different optimization goals; the most relevant variant \emph{minimizes the maximal load} of a voter. 

In particular, \emph{\maxP} is the rule that returns winner sets corresponding to load distributions minimizing the maximal voter load. 
And \emph{\seqP} is a sequential (greedy) version of \maxP; it selects candidates iteratively, in each round adding a candidate to the committee such that the new maximal voter load is as small as possible. 

\subsubsection{Load Distributions}

Given an election $\sigma = (N, C, \mathcal{A}, k)$ and a subset $D \subseteq C$ of candidates, %
a \emph{load distribution for $D$ given $\sigma$} is a two-dimensional array ${\bf \ell}= (\ell_{i,c})_{i \in N, c \in D}$ satisfying
\begin{eqnarray*}
0 \leq \ell_{i,c} \leq 1 & & \text{for all $i \in N$ and $c \in D$,} \\
\ell_{i,c}= 0 & & \text{for all $i \in N$ and $c \in D \setminus A_i$, and} \\
\sum_{i \in N} \ell_{i,c} = 1 & & \text{for all $c \in D$}. 
\end{eqnarray*}
\newcommand{\ld}{\mathcal{L}}
\newcommand{\ldo}{\mathcal{L}^{\textrm{opt}}}
We let $\ld_{\sigma,D}$ denote the set of all load distributions for $D$ given $\sigma$. For a load distribution $\ell \in \ld$, the \emph{total load of voter $i$ under $\ell$}, denoted $\ell_i$, is given by $\ell_i= \sum_{c \in D} \ell_{i,c}$.
Note that $\sum_{i \in N} \ell_i = |D|$ for all $\ell \in \ld_{\sigma,D}$.

Finally, a load distribution is called \emph{optimal} for given $\sigma$ and $D$ if the maximal total voter load $\max_{i \in N} \ell_i$ is as small as possible. $\ldo_{\sigma,D}$ denotes the set of all optimal load distribution functions.  

We are now going to establishing a close connection between load distributions and support distribution functions.\footnote{Throughout this section we assume that each
  candidate in $D$ is approved by some voter in $N$; otherwise, load
  distributions cannot be defined.}

\begin{lemma}\label{lem:s-l}
Let $\sigma = (N, C, \mathcal{A}, k)$ be an \abe and $D \subseteq C$ a
subset of candidates. Then, the following statements
hold.
\begin{enumerate}
	\item For every support distribution function $f \in \sd_{\sigma,D}$, there is a load distribution $\ell^f \in \ld_{\sigma,D}$ such that 
	 \[ \max_{i \in N} \ell^f_i \le \frac{1}{\min_{c \in D} \supp_f(c)}.\]
	\item For every load distribution $\ell \in \ld_{\sigma,D}$, there is a support distribution function $f^\ell \in \sd_{\sigma,D}$ such that 
	\[ \min_{c \in D} \supp_{f^\ell}(c) \ge \frac{1}{\max_{i \in N} \ell_i}.\]
\end{enumerate}
\end{lemma}

\begin{proof}
For a given a support distribution function $f \in \sd_{\sigma,D}$,
define the load distribution $\ell^f \in \ld_{\sigma,D}$ by setting
$\ell^f_{i,c} = \frac{f(i,c)}{\supp_f(c)}$ for each $i \in N$ and $c \in
D$.\footnote{If for some candidate $c$ it is $\supp_f(c)= 0$ the first part of the lemma trivially holds.}
It follows that the total load of a voter is upper bounded by
$\frac{1}{\supp_f(c^*)}$, where $c^*$ is a candidate with minimal
support (we recall that $\sum_c f(i,c)= 1$ for each voter $i$ such
that $A_i \cap D \ne \emptyset$).

For a given load distribution $\ell \in \ld_{\sigma,D}$, define a
support distribution function $f^\ell \in \sd_{\sigma,D}$ by setting $f^\ell(i,c) = \frac{\ell_{i,c}}{\ell_i}$ for each voter $i \in
N$ such that $\ell_i > 0$.  That is, the support for a candidate is
proportional to the load received from that candidate, scaled such
that the total support by the voter is 1.  It follows that the minimal
support of a candidate is lower bounded by $\frac{1}{\ell_{i^*}}$,
where $i^*$ is a voter with maximal load.  To see this, let $i^*$
denote a voter with maximal load and let $c \in
D$. Then, \[\supp_{f^\ell}(c) = \sum_{i \in N} f^\ell(i,c) \geq \sum_{i
  \in N: \ell_i > 0} \frac{\ell_{i,c}}{\ell_i} \ge \frac{1}{\ell_{i^*}} \sum_{i
  \in N} \ell_{i,c} = \frac{1}{\ell_{i^*}}.\]
\end{proof}

\subsubsection{\phrag's Optimal Rule}

Lemma~\ref{lem:s-l} has particularly interesting implications for load distributions and support distribution functions that are optimal: 
The construction used in the proof of Lemma~\ref{lem:s-l} establishes a one-to-one relationship between elements of $\ldo_{\sigma,D}$ and elements of $\sdo_{\sigma,D}$. Therefore, the objective of minimizing the maximal voter load is equivalent to the objective of maximizing the minimal support.  
As a consequence, \maxP (the method that globally minimizes the maximal voter load) is identical to the rule that globally maximizes the minimal support.\footnote{The latter method has been referred to as \emph{optimal open D'Hondt (OODH)} in earlier versions of this manuscript.}

\begin{theorem}
	Let $\sigma = (N, C, \mathcal{A}, k)$ be an \abe. Then,
	$\textrm{\maxP}(\sigma) = {\arg \max}_{W \subseteq C, |W| = k} \mm(\sigma, W)$.
\end{theorem}

Since it is NP-hard to compute winners under \maxP
\citep{brill:phragmen}, the same is true for finding a set of
candidates maximizing the maximin support. \citet{brill:phragmen}
proved that \maxP satisfies PJR (when combined with an appropriate
tie-breaking rule) but not EJR. With respect to monotonicity axioms,
\citet{mora2015eleccions} proved that \maxP fails house monotononicity
and~\citet{2017arXiv171004246S} have recently extended previous
results by~\citet{phragmen:pOpt} showing that \maxP satisfies weak
support monotonicity but fails strong support monotonicity.

\subsubsection{\phrag's Sequential Rule}

There is also a close relationship between the \ms method ($\mms$) and \phrag's \emph{sequential} rule (\seqP). Both $\mms$ and \seqP construct the set of winners by iteratively adding candidates: $\mms$ chooses candidates such that the minimal support of the new set is maximized; \seqP chooses candidates such that the maximal voter load incurred by the new set is minimized.    
However, there is a subtle difference between the two methods concerning the \emph{redistribution} of support/load.
Under $\mms$, support distributed to candidates in earlier rounds can be freely redistributed when looking for maximin support distributions for the new set of candidates. This is not the case for the loads under \seqP, however: once a voter is assigned some load from some candidate, this load is ``frozen'' and will always stay with the voter. As a consequence, the two methods might give different results.

\begin{example}
  Consider again the \abe $\sigma_4$ from Example~\ref{ex:EJR}. We
  recall that $\mms(\sigma_4) = \{c_1, a_1, a_2, a_3\}$. In contrast,
  it can be easily shown that \seqP selects (in this order) $\{c_1,
  c_2, c_3, a_1\}$.
\end{example}  

It is straightforward to check that \seqP can be computed in polynomial time. With respect to the axiomatic properties considered in this paper, \seqP is indistinguishable from the \ms method: \seqP
satisfies house monotonicity by definition; it satisfies PJR but fails EJR \citep{brill:phragmen}; and results
by~\citet{phragmen:pOpt},~\citet{mora2015eleccions},
and~\citet{2016arXiv161108826J} imply that \seqP satisfies weak support
monotonicity but violates strong support monotonicity.

\subsection{Thiele's Rules}

Thiele's rules are based on a score optimization problem \citep{thiele:pav_rav}. For a given \abe $\sigma=(N, C, \mathcal{A}, k)$, the goal is to find a winner set $W$ with $|W|=k$ maximizing $s(W) = \sum_{i \in N} s(i,W)$, where $s(i,W)$ is defined by $s(i,W) = \sum_{j=1}^{|A_i \cap W|} \frac{1}{j}$. 
\citet{thiele:pav_rav} proved that his methods are extensions of the D'Hondt method (see also \citep{2016arXiv161108826J,brillmultiwinner}).


Thiele's \emph{optimal rule}, often referred to as \emph{Proportional Approval Voting (PAV)} \citep{kilgour10}, outputs a set $W$ maximizing $s(W)$. 
This rule satisfies EJR (and thus PJR) \citep{aziz:scw} and is NP-hard to compute \citep{aziz2014computational,skowron2016finding}. It was already known by~\citet{thiele:pav_rav} that PAV fails house monotonicity.~\citet{2017arXiv171004246S} proved that PAV satisfies strong support monotonicity with population increase (in fact, PAV is the only rule that is known to satisfy this axiom and EJR or PJR) but only weak support monotonicity without population increase. 

Thiele's \emph{sequential rule} (sometimes referred to as \emph{Reweighted Approval Voting (RAV)}, \emph{sequential PAV}, or \emph{Thiele's addition method}), is a greedy heuristic for the score optimization problem defined above.
The rule starts with $W= \emptyset$ and iteratively adds candidates $c$ maximizing the score $s(W \cup \{c\})$.
It is straightforward to show that Thiele's sequential rule can be computed in
  polynomial time. 
The rule satisfies house monotonicity by definition. Furthermore, it satisfies weak support monotonicity but not strong support monotonicity~\citep{2017arXiv171004246S}, and it
fails PJR \citep{aziz:scw}.\footnote{Thiele's sequential method also violates the weaker property \emph{justified representation}~\citep{aziz:scw,pjr-aaai}.}

\section{Conclusions}
\label{sec:concl}

In this paper we have proposed the \ms method, a novel extension of the D'Hondt method to approval-based multiwinner elections. The principle on which this voting rule is based is that the support of the least supported winner should be maximized.
 
 We have established that the \ms method can be computed efficiently and satisfies a number of appealing axiomatic properties, including house monotonicity, weak support monotonicity, and proportional justified representation. We have also shown that the rule admits two equivalent formulations: in each iteration, selecting the candidate with the highest support (using an optimal support distribution function) is equivalent to selecting the candidate that maximizes the support of the least supported candidate. This can be seen as selecting, at the same time, the ``best'' candidate and the ``best'' set of candidates (from those that can be obtained by adding a new candidate to the set of previously chosen winners). We believe that this is a nice feature of the \ms method. 

  We have also established a close relationship between the \ms method and \phrag's rules. This novel connection allows us to formulate \phrag's optimal rule as a support maximization (rather than a load minimization) problem, and to view the \ms method as a tractable approximation of \phrag's (intractable) optimal rule. It would be very interesting to further illuminate the differences between the \ms method and \phrag's sequential rule.     %

Other possible lines of future work include axiomatic characterizations of the \ms method (and other approval-based multiwinner rules), as well as the development of extensions to the ranked ballot setting.

\section{Acknowledgements}

We would like to thank the anonymous reviewers for
their helpful comments, which have helped to improve the paper significantly. 
A previous version of this paper has circulated under the title ``Fully Open Extensions to the D'Hondt Method'' \citep{2016arXiv160905370S} and was presented at the 13th Meeting of the Society for Social Choice and Welfare (Lund, June 2016). In this earlier version, the \ms method was referred to as the \emph{Open D'Hondt (ODH)} method.

This work was supported in part by the Spanish Mi\-nis\-terio de
Econom\'{\i}a y Competitividad (project AUDACity TIN2016-77158-C4-1-R) and by a Feodor Lynen return fellowship of the Alexander von Humboldt Foundation.

%


\bibliographystyle{plainnat}
\bibliography{dhondt,abb,markus}

\newpage

\appendix

\section{Proof of Theorem \ref{th:lastIsLeast}}
\label{app:proof}

We employ linear programming duality theory (see, e.g., \citep{Chva83a}).
Let $\sigma$ be an \abe and $D \subseteq C$ a nonempty subset of candidates. We have seen in Section~\ref{sec:odh} that $\mm(\sigma,D)$ can be computed with the following linear program.
\begin{alignat}{3}
 & \text{maximize} &s & \nonumber\\
 & \text{subject to} \quad& 
    \sum_{\mathclap{{i \in N: c \in A_i \cap D}}} f(i,c) & \ge s, &\text{for all $c \in D$} \label{con:c}\\
	&& \sum_{\mathclap{{c \in A_i  \cap D}}} f(i,c) & = 1, \quad &\text{for all $i \in N$ with $A_i \cap D \neq \emptyset$} \label{con:i}\\
	&& f(i,c) &\ge 0, &\text{for all $i \in N$ and $c \in D$} \nonumber
\end{alignat}
We now consider the \emph{dual} of this linear program. For every inequality constraint of the form (\ref{con:c}), there is an associated non-negative dual variable~$y_c$ ($c \in D$), and for every equality constraint of the form (\ref{con:i}), there is an unrestricted dual variable~$z_i$ ($i \in N$ and $A_i \cap D \neq \emptyset$).
The dual looks as follows.
\begin{alignat}{3}
 & \text{minimize} \quad &\sum_{\mathclap{i \in N: A_i \cap D \neq \emptyset}} \, z_i & \nonumber \\ 
 & \text{subject to} \quad& \sum_{\mathclap{c \in D}} \, y_c &=1 & \label{con:one}\\
	&& z_i &\ge y_c, \quad &\text{for all $i \in N$ and $c \in A_i \cap D$} \label{con:z} \\[2ex]
	&& y_c &\ge 0, &\text{for all $c \in D$} \nonumber
\end{alignat}
Let $(s^*,f^*)$ be an optimal solution for the primal linear program and
$(y^*,z^*)$ an optimal solution for the dual linear program. 
Then, 
\[s^* \quad=\quad \sum_{\mathclap{i \in N: A_i \cap D \neq \emptyset}} z^*_i \quad=\quad \mm(\sigma,D)\text.\]

Moreover, from \emph{complementary slackness} we get that the implication
\begin{equation}\label{eq:cs}
	\supp_{f^*}(c) > s^*  \quad \Rightarrow \quad  y_c^* = 0 
\end{equation}
holds for all candidates $c \in D$. That is, if the support of a candidate $c$ under an optimal support distribution function $f^*$ exceeds $\mm(\sigma,D)$, then the dual variable $y_c$ 
(associated with the primal constraint ${\sum_i f(i,c) \ge s}$) 
equals zero in the optimal solution $(y^*,z^*)$ for the dual.

We first prove a lemma that relates maximin support values of different sets. In particular, it states that candidates receiving more then the minimal support (under an optimal support distribution function) can be removed without affecting the maximin value of the set. 
 
\begin{lemma}
\label{lem:kernel}
Let $\sigma = {(N, C, \mathcal{A}, k)}$ be an \abe, 
$D \subseteq C$ a nonempty subset of candidates, 
and $f \in \sdo_{\sigma,D}$ an optimal support distribution function. 
If there exists a candidate $\hat{c} \in D$ such that $\supp_f(\hat{c}) > \mm(\sigma, D)$, 
then $\mm(\sigma, D) = \mm(\sigma, D \setminus \{\hat{c}\})$.
\end{lemma}

\begin{proof}
	It is easy to verify that $\mm(\sigma, D) \le \mm(\sigma, D \setminus \{\hat{c}\})$. We will now show that $\mm(\sigma, D) \ge \mm(\sigma, D \setminus \{\hat{c}\})$ also holds. 
	
	Consider the primal and dual linear programs corresponding to the computation of $\mm(\sigma,D)$ together with optimal solutions $(s^*,f^*)$ and $(y^*,z^*)$, where $f^*=f$ and $s^*=\mm(\sigma,D)$. 
	Since $\supp_f(c) > \mm(\sigma, D)$ holds by assumption, (\ref{eq:cs}) implies that $y_{\hat{c}}^*=0$.
	Furthermore, if there exist voters $i \in N$ with $A_i \cap D = \{\hat{c}\}$, then $z_i^*=0$ for all such voters (because variable $z_i$ only appears in the single constraint $z_i\ge y_{\hat{c}}$).  
	
	Now consider the dual linear program corresponding to the computation of $\mm(\sigma,D \setminus \{\hat{c}\})$. This linear program has a variable $y_c$ for each $c \in D \setminus \{\hat{c}\}$ and a variable $z_i$ for each $i \in N$ with $A_i \cap (D \setminus \{\hat{c}\}) \neq \emptyset$. We are going to construct a feasible solution $(\hat{y},\hat{z})$ for this linear program by restricting $(y^*,z^*)$ to the smaller domain. 

	For each $c \in D \setminus \{\hat{c}\}$, let $\hat{y}_c = y^*_c$.  Moreover, for each $i \in N$ with $A_i \cap (D \setminus \{\hat{c}\}) \neq \emptyset$, let $\hat{z}_i = z^*_i$. The solution $(\hat{y},\hat{z})$ is feasible for the linear program in question because $y_{\hat{c}}^*=0$ and thus 
	\[\sum_{c \in D \setminus \{\hat{c}\}} \hat{y}_c \quad=\quad \sum_{c \in D} y^*_c \quad=\quad 1 \text.\]
And the objective function value of the solution $(\hat{y},\hat{z})$ is equal to that of the solution $(y^*,z^*)$ in the original dual because
	\[
	            \sum_{\mathclap{\substack{i \in N: \\ A_i \cap (D \setminus\{\hat{c}\}) \neq \emptyset}}} \hat{z}_i 
	\quad=\quad \sum_{\mathclap{\substack{i \in N: \\ A_i \cap (D \setminus\{\hat{c}\}) \neq \emptyset}}} z^*_i 
	\quad=\quad \sum_{\mathclap{\substack{i \in N: \\ A_i \cap (D \setminus\{\hat{c}\}) \neq \emptyset}}} z^*_i 
	\quad+\quad \sum_{\mathclap{\substack{i \in N: \\ A_i \cap D = \{\hat{c}\}}}} z^*_i 
	\quad=\quad \sum_{\mathclap{\substack{i \in N: \\ A_i \cap D \neq \emptyset}}} z^*_i
	\quad=\quad s^* \text.\]

It follows that the objective function value of the dual linear program for $D \setminus \{\hat{c}\}$ is less than or equal to the objective function value $s^*$ of the dual linear program for $D$.\footnote{Recall that the duals are minimization problems.} In other words, $\mm(\sigma,D \setminus \{\hat{c}\}) \le s^* =  \mm(\sigma,D)$. 
\end{proof}

We are now ready to prove Theorem \ref{th:lastIsLeast}.

\setcounter{theorem}{0}
\begin{theorem}
Let $\sigma=(N, C, \mathcal{A}, k)$ be an \abe. The following holds for each
$j \in \{0, \ldots, k-1\}$.

Let {$W^j$} denote the set of the first $j$ candidates chosen by the \ms method when applied to $\sigma$.
Then, for each candidate $c \in C \setminus W^j$ and for each
optimal support distribution function
$f_c \in \sdo_{\sigma, (W^j \cup \{c\})}$,
\[\supp_{f_c}(c) = \mm(\sigma, W^j \cup \{c\})\text.\]
\end{theorem}

\begin{proof}[Proof of Theorem \ref{th:lastIsLeast}]
The proof is by induction on $j$.
For $j= 0$, the statement clearly holds because 
$W^0 \cup \{c\}= \{c\}$ and there is a unique optimal support
  distribution function $f_c \in \sdo_{\sigma,\{c\}}$ that furthermore satisfies $\supp_{f_c}(c) = \mm(\sigma, \{c\})$.

For the inductive step, let us assume that the statement holds for $j=m$. We show that it also holds for $j= m+1$. 

Suppose for contradiction that this is not the case. Then, there exist a candidate 
$c \in C - W^{m+1}$ and an optimal support distribution function $f_c \in \sdo_{\sigma,
    W^{m+1} \cup \{c\}}$ such that

\begin{equation}
\supp_{f_c}(c) >
\mm(\sigma, W^{m+1} \cup \{c\}). 
\label{eq:th_lil_1}
\end{equation}

By Lemma~\ref{lem:kernel}, this implies that

\begin{equation}
\label{eq:th_lil_2}
\mm(\sigma, W^{m+1} \cup \{c\}) =
\mm(\sigma, W^{m+1}).
\end{equation}

Let {$c_{m+1}$} be the {$(m+1)$st}
candidate chosen by $\mms$ for election $\sigma$. Thus,
{$W^{m+1}= W^m \cup \{c_{m+1}\}$}. 
Let $g$ be a support distribution function on $W^m \cup \{c\}$
such that
\begin{equation}
\label{eq:th_lil_3}
  \supp_g(c') \geq \supp_{f_c}(c') \quad \text{for all $c'$ in $W^m \cup \{c\}$.}
\end{equation}
Such a function can easily be constructed by considering $f_c$ and redistributing support that is assigned to candidate $c_{m+1}$.

We now distinguish two cases: either the function {$g$}
maximizes the support for the least supported candidate in
{$W^m \cup \{c\}$}, or it does not.

\paragraph{Case 1} $g \in \sdo_{\sigma, W^m \cup \{c\}}$. In this case,

\begin{equation}
\mm(\sigma, W^m \cup \{c\}) = \min_{c'
  \in W^m \cup \{c\}} \supp_g(c'),
\end{equation}

and thus, by the induction hypothesis,

\begin{equation}
\label{eq:th_lil_4}
\mm(\sigma, W^m \cup \{c\}) = \supp_g(c).
\end{equation} 

By combining~(\ref{eq:th_lil_1}), (\ref{eq:th_lil_2}),
(\ref{eq:th_lil_3}), and~(\ref{eq:th_lil_4}), we have

\begin{eqnarray*}
\mm(\sigma, W^m \cup \{c\}) &=& \supp_g(c) \geq \supp_{f_c}(c) \\
&>& \mm(\sigma, W^{m+1} \cup \{c\}) \\ 
&=& \mm(\sigma, W^m \cup \{c_{m+1}\}).
\end{eqnarray*}

However, this is a contradiction because it implies that candidate $c$ should have been
selected instead of {$c_{m+1}$} at iteration {$(m+1)$}.

\paragraph{Case 2} $g \in \sd_{\sigma, W^m \cup \{c\}} \setminus \sdo_{\sigma, W^m \cup \{c\}}$. In this case,

\begin{equation}
\label{eq:th_lil_5}
\mm(\sigma, W^m \cup \{c\}) > \min_{c'
  \in (W^m \cup \{c\})} \supp_g(c').
\end{equation}

Furthermore, since $f_c \in \sdo_{\sigma,W^{m+1} \cup \{c\}}$, we have
\begin{equation}
\label{eq:th_lil_6}
\min_{c' \in (W^{m+1} \cup \{c\})} \supp_{f_c}(c') = \mm(\sigma, W^{m+1} \cup \{c\}) \text.
\end{equation} 

By combining~(\ref{eq:th_lil_2}), (\ref{eq:th_lil_3}), (\ref{eq:th_lil_5})
and~(\ref{eq:th_lil_6}), we have

\begin{eqnarray*}
\mm(\sigma, W^m \cup \{c\}) &>& {\displaystyle
  \min_{c' \in (W^m \cup \{c\})} \supp_g(c')}
\geq {\displaystyle \min_{c' \in (W^{m+1} \cup
    \{c\})} \supp_{f_c}(c')} \\ 
&=& \mm(\sigma, W^{m+1} \cup \{c\}) \\
&=& \mm(\sigma, W^m \cup \{c_{m+1}\}).
\end{eqnarray*}

Again, this is a contradiction because it implies that candidate $c$ should have been
selected instead of {$c_{m+1}$} at iteration {$(m+1)$}.  
\end{proof}

\end{document}